\documentclass[aps,prl,groupedaddress,reprint,showpacs]{revtex4-1}
\usepackage{graphicx}
\usepackage[right]{eurosym}
\newcommand{\avg}[1]{\ensuremath{\left\langle #1 \right\rangle}}

\begin{document}
\title{On house renovation and coauthoring  \\(with a little excursus on the Holy Grail of bibliometrics)}
\author{Roberto Piazza}
\affiliation{Department CMIC ``Giulio Natta'', Politecnico di
Milano, via Ponzio 34/3, 20133 Milano, Italy}
\altaffiliation[Present address: ] {Cavendish Laboratory,
Cambridge, UK}
\date{\today}

\begin{abstract}
More than a paper, this is just a little divertissement about
coauthoring, the Hirsch $h$-index, and bibliometric evaluation in
general. Without pretending to yield any general conclusions, what
I found rummaging through the physics literature made me think
quite a bit. I hope the same will happen to my readers, even it
they will likely be much less than 25, which is the audience one
of the greatest Italian writers (whom, is left to the reader to
single out) addresses to.

\end{abstract}

\maketitle
 Suppose that your house needs some restoration, and that you call
a master mason asking for an estimate. If the mason replies at
once that he will quote \EUR{1000} for himself, plus \EUR{500} for
each helper apprentice, you will likely be puzzled, if not
annoyed. Surely you have good reasons to complain, reasoning that
the job you ask for should be remunerated with a fixed amount,
irrespective of the number of labourers it requires. Yet, this is
not a criterium that we usually apply when evaluating the CV of an
applicant for an academic position or for a grant. We may examine
the number of papers the applicant has made, where they have been
published, or how many citations they have obtained. More
recently, we would surely check the Hirsch $h$-index
\cite{Hirsch}, or exploit more sophisticated indicators. Rarely we
look for the extent of coauthoring: a good paper is a good paper
and, in terms of the applicant prestige, it is often regarded to
be equally worth regardless it is signed by one, five, or two
hundreds coauthors. Possibly, if the applicant is the first
author, who presumably made the hard job, or the last one, usually
the lab ``master mason'', you may grant her or him an additional
bonus. But that's all. After all, recovering quantitative
information of this kind from search services like ISI or Scopus,
even something simple as the average number of coauthors per
paper, is not immediate (just try!).

Suppose however that the mason refutes your argument by claiming
that the more people do the job, the better it comes out. You may
be skeptical, but you will not easily come out with general
abstract arguments for or against such a claim. Like a cosmologist
who has a single Universe to investigate (if she or he is an
experimentalist, at least), you have just this house to test, and
relying on repetitive trials is out of question (besides
expensive). Grounding discussions about coauthoring on abstract
arguments is conversely not uncommon in the scientific community,
at least in my native country. Some colleagues argue that, yes,
discouraging excessive coauthoring is probably sensible, but that
a penalty consisting in simply dividing the citations of a given
paper by the number $N$ of authors  is probably excessive. So they
suggest using diverse sublinear functional forms of scaling, such
as dividing by $\sqrt{N}$, usually on the basis of some kind of
\emph{a priori} reasoning. Some others (mostly experimentalists),
however, reply that being able to build up a collaboration network
is a virtue that should be acknowledged, hence no scaling should
be applied if $N$ is still moderately large, say, smaller than 5
or 10. When questioned, certain physicists - for some obscure
reason, usually high energy experimentalists - even let the matter
drop at once, branding talks of this kind as ``absurd''.

\begin{figure}[b]
\centering
  \includegraphics[width=\columnwidth]{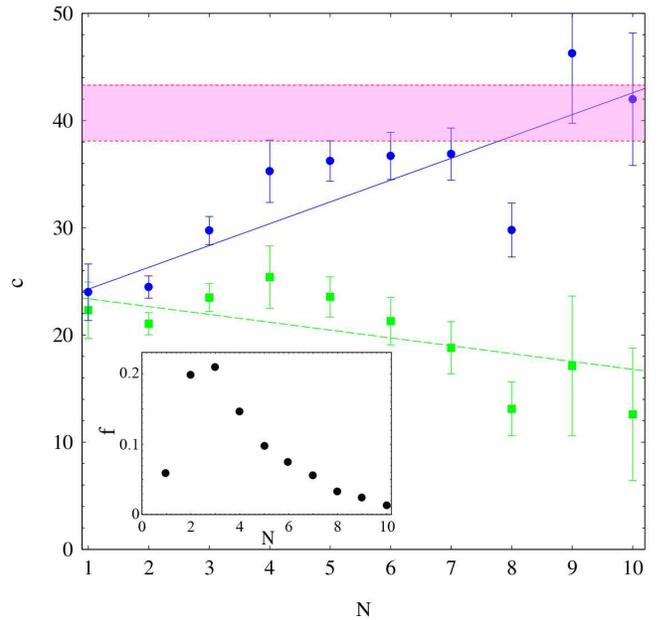}
  \caption{Average number of citations $\bar{c}$ versus the number $N$ of
  coauthors  at the end of 2012 for the manuscripts published in Phys. Rev. Lett. in 2007
with $N\le 10$ (blue dots). Data are obtained from a set of about
3400 records, with the distribution shown in the inset. The full
line is a linear fit with slope
  $(0.08\pm0.02)N$. The purple band shows the number of citations (within $\pm 1 \sigma$) of the papers with $N>10$, which are about $8\%$ of the total.
  When self-citations are tentatively removed by rescaling $\bar{c}$ by a factor
  $(1+0.07N)^{-1}$, the corrected data point (green squares) show no significant change (or even a slight decrease)
  with $N$.\label{f1}}
\end{figure}
The fact is,  at variance with the former case, we \emph{do} have
a sensible, albeit not perfect way to quantify how much
coauthoring impacts on the recognition of a publication by looking
at the total number of citations it has received after some years.
I have then considered the number of citations in the first 6
years, according to ISI Web of Knowledge (WoK), by all manuscripts
published in Physical Review Letters \footnote{Admittedly,
scientists publishing in PRL are already a rather selected group:
those physicists that can boast many papers in this prestigious
journal, still a reference in our community, is probably a minor
fraction. Nevertheless, the latter arguably includes also those
young scientists we may wish to consider for a position.} in 2007
(about 3700 records, including comments but not replies and
corrections). I have then sorted these papers in groups on the
basis of the number of authors, and evaluated the average and
standard deviation of the number of citations $c$ for each group.
A first striking evidence from the results, shown in
Fig.~\ref{f1}, is that $c$ grows by a mere factor of two when $N$
increases from 1 to 10, namely, just a little more than 8\% for
each additional author. Equally surprising is that, as clearly
evidenced by the purple band in Fig.~\ref{f1}, very large
collaborations do not seem to yield, on the average, a much
greater impact on the scientific community. In other words, if we
``reward'' each author just on the basis on the total number of
citations he/she has obtained, we are likely to make a big gift to
those masons used to work in large groups. Nevertheless, a
moderate increase with $N$ of the ``acknowledged value'' of a
publication seems to be present.

At least, if we neglect self-citations. Quantifying the latter for
each single record is a hard task, and the WoK is surely not of
great help. Just to get a rough figure, I then simply considered
the average fraction of self-citation for those authors (about
150) of the 5\% most cited papers who have got an ISI Author
Identifier , which turns out to be $0.07\pm 0.01$. If we then
assume that each of the coauthors contributes to the total number
of a citations of a given paper with 7\% of self-citations, we may
think of subtracting out this ``spurious'' contribution by
substituting $c \rightarrow (1-0.07N)c$. This is of course
questionable, since several papers have been probably cited by
more than one coauthor, hence the contribution of self-citations
is likely to be overestimated. Nevertheless, the result is rather
impressive, for the net data obtained this way (squares in Fig.
\ref{f1}) even show a  slight apparent \emph{decrease} with $N$.
Summing up, I am prone to conclude that the ``merit'' of a
scientific publication, as judged by the number of citations it
obtains, does not basically depend in $N$. Hence, in the absence
of further information on the role played by each author (of the
kind provided for instance in several biological or medical
journals), credit should be shared in equal parts by all
coauthors.

In bibliometric assessments, taking into account these ``profit
sharing'' considerations in detail might be hard. A crude but
reasonable approach could simply be rescaling the total number of
the citations of a scientist by the \emph{average} number of
coauthors of her/his papers - an information, however, which is
not readily obtained from search services - or, in the case of the
$h$-index, by the average number of authors of her/his $h$ most
cited papers. A brief excursus on latter, however, may be useful.
Because it is so easy to evaluate, but more than that because of
its statistical robustness, the Hirsch index has rapidly ascended
the throne of bibliometrics as a single number summarizing the
success of a scientist. I must admit that, living in a country
where quantitative evaluation of quality has always been seen with
suspicion (and often, when made, easily circumvented
\footnote{According to the traditional Italian saying ``fatta la
legge, gabbato lo santo'', which roughly means ``once the rule is
established, the saint is duped''.}), I have been a fan, or almost
a zealot of this brilliant, straightforward approach since it was
originally proposed. Yet, how much additional information does the
Hirsch index \emph{really} convey? We may reasonably expect $h$ to
scale with order $\sqrt{c}$~\footnote{At least, $\sqrt c$ is
obviously an upper bound for $h$.}. But is there any relation
between $h$ and the total number of papers an author has
published?

\begin{figure}[b]
\centering
  \includegraphics[width=\columnwidth]{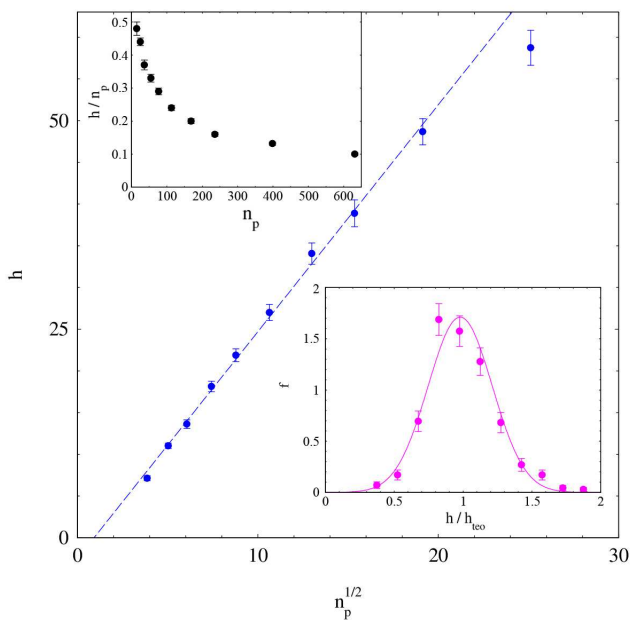}
  \caption{Main figure: Average Hirsch index $h$ as a function of the square root of the number of published papers $n_p$,
  for a set of 470 scientists co-authoring the top 10\% cited papers published by Phys. Rev. Lett.
  in 2012, fitted as \mbox{$h_{teo} =(2.72\pm 0.05)n_P^{1/2}-(2.5\pm0.5)$}.
  The dependence on $n_p$ of the ``success ratio'' $h/n_p$ is shown in the upper inset. The lower inset gives the
  relative frequency distribution of the quantity  $h/h_{teo}$ for the whole set of investigated authors, fitted with a
  gaussian of standard deviation $\sigma =0.23$.\label{f2}}
\end{figure}

To this aim, I have considered the 10\% most cited papers
published in PRL last year (2012), (manually) examining the
individual citation reports of all those authors (470 in total)
who appear to have an ISI Author ID. The upper inset in
Fig.~\ref{f2} shows that, as it can be reasonably expected, the
ratio $h/n_p$ of the number of papers that contribute to the
$h$-index to the total number of papers $n_p$ an author publishes
(which we could consider as a kind of ``success ratio'') rapidly
decreases with $n_p$. Actually, the main body of Fig.~\ref{f2}
shows that $h$ is quite well fitted by a linear dependence on
$\sqrt{n_p}$, except for $n_p \gtrsim 400$, where some saturation
may be present. What is really surprising is the very limited
dispersion of the data around the mean. As a matter of fact, the
ratio between the actual value of $h(n_p)$ for the individual
authors and the value one gets from the fit to the data has an
approximately gaussian distribution, with a standard deviation
$\sigma =0.23$.

In simple words, this means the following: tell me the total
number of papers you have published, and I'll predict your
$h$-index within $20-30\%$ accuracy. More seriously, this result
cast doubts on the amount of novel information  the $h$-index
carries \emph{per se}, besides a simple reshuffling of a basic and
rather trivial information about an the total scientific
productivity of an author. In fact, provided that these general
observations are confirmed by testing a much larger and varied
sample besides the limited and rather selected one I have
considered, surely not representative of the whole population of
physicists~\footnote{Such a test, which could be easily made by
ISI or Scopus, would likely yield a larger dispersion of $h$
around $h_{teo}$.}, a more meaningful bibliometric parameter would
actually be the \emph{deviation} $\delta h = h/h_{teo}- 1$.

\begin{figure}[h]
\centering
  \includegraphics[width=\columnwidth]{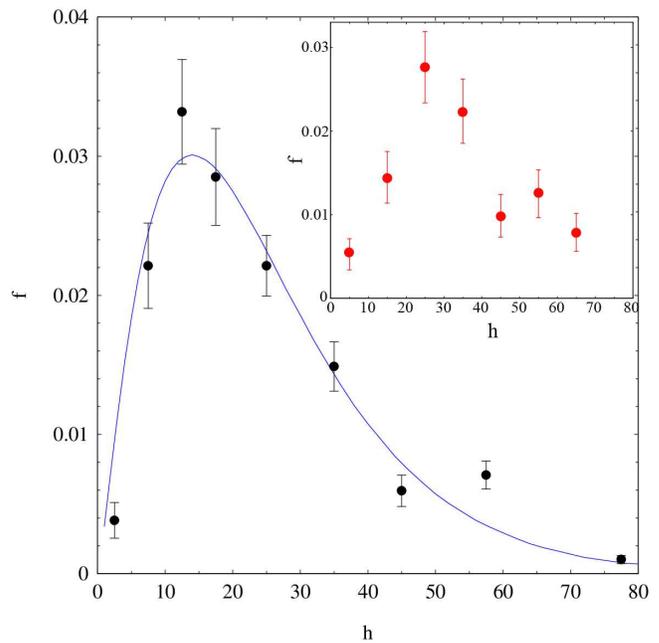}
  \caption{Frequency distribution of the $h$-index for the author set shown in Fig.~\ref{f2},
  fitted with a Gamma distribution
  $\Gamma(h;\alpha =2.27, \beta =11.0)$. The inset displays the frequency distribution
  for the subset of about 150
  authors belonging to large collaborations. \label{f3}}
\end{figure}
The combination of the basic independence of the value of a
scientific paper from $N$ with the former tight statistical
relation between $n_p$ and $h$ would, if confirmed, be
particularly significant for those physicists belonging to large
collaborations such as Atlas,  LHCb, CDF, and so on. The main body
of fig.~\ref{f3} shows that the frequency distribution of the
$h$-index for those authors considered in Fig.~\ref{f2}, which has
an average value $\bar{h} \simeq 27$  and a relative standard
deviation $\sigma_h/\bar{h} \simeq 0.63$ is, as may be expected,
considerably skewed. The distribution is indeed approximately
fitted by a Gamma PDF with an expectation value $\avg{h} \simeq
25$ and a much lower mode $h_{max} \simeq 14$~\footnote{I leave it
to the reader to brood over the origin of this peculiar
distribution, suggesting that sampling over the PRL authors is
basically a random Poisson process. For the aims of this paper, it
is sufficient to note that the distribution covers a wide spectrum
of values for the $h$-index, fairly representing both young
postdocs and ageing professors like me.}. However, the inset shows
that the same distribution, when restricted only to those authors
belonging to large collaboration groups, has a rather different
shape, being almost symmetric, with a larger average value
$\bar{h} \simeq 34$ but a lower relative standard deviation
$\sigma_h/ \bar{h} \simeq 0.47$. These means that these authors,
besides being inclined to publish more (recall, however than, on
the average, collaboration papers are \emph{not} cited much more
than papers with a few authors), and form a more homogeneous group
in term of their overall ``scientific success''. Note that, in
this restricted distribution, low values of the $h$-index are
consistently less represented. hence, either young scientists are
less frequently included in the authors' list or, more likely,
belonging to large collaboration groups rewards young physicists
by allowing them to coauthor so many papers that their
bibliometric parameters rapidly rise to values which are typical
of more mature scientists. In any case, the relative homogeneity
of the population, together with the limited credit that,
according to Fig.~\ref{f1}, should be given to a single individual
for the acknowledgement of works made by large groups, makes the
$h$-index a rather poor evaluation parameter to differentiate
among young high-energy or nuclear physicists.

As I mentioned in the abstract, this little \emph{divertissement}
should not be taken too seriously, for any sound conclusions must
be corroborated by a much more extensive and rigorous statistical
analysis. The former observations, however, lead me to two
considerations. For what concerns myself, in the future I would
not like to take part in committees where hiring or funding of
young scientists is made only on bibliometric bases, renouncing to
the pleasure of interviewing, even shortly, the candidates. For
what concerns my fellow countrymen, the warning is that no
bibliometric approach to hiring and promoting, however refined,
will ever ensure a real improvement of our academic institutions,
unless there are ultimate motivations \emph{to long} for
scientific quality. And this, in a country where competition
between universities is still seen with suspicion - ``rating, but
not ranking'' is a basic recommendation of our National University
Council (CUN) \footnote{CUN official declaration to the Ministry
on the Evaluation of the Quality of Research (VQR), 16 July 2013}-
is far from being a priori ensured.

Finally, let me thank Pietro Cicuta for having invited me here in
Cambridge, where (besides doing some real work), I managed to find
some time for idling with these trifles. I have also took
pleasure from discussing these issues with Wilson Poon, a
scientist well on the right (in both senses) side of the gaussian
in Fig. \ref{f2}.
%

\end{document}